\documentclass[prb,showpaces,preprintnumbers,amsmath,amssymb,twocolumn,superscriptaddress,longbibliography]{revtex4-1}
\usepackage{graphicx}
\begin{document}

\title{Unconventional superconductivity in the cage type compound Sc$_5$Rh$_6$Sn$_{18}$}
\author{A. Bhattacharyya}
\email{amitava.bhattacharyya@rkmvu.ac.in} 
\affiliation{ISIS Facility, Rutherford Appleton Laboratory, Chilton, Didcot Oxon, OX11 0QX, United Kingdom} 
\affiliation{Highly Correlated Matter Research Group, Physics Department, University of Johannesburg, PO Box 524, Auckland Park 2006, South Africa}
\address{Department of Physics, Ramakrishna Mission Vivekananda Educational and Research Institute, Belur Math, Howrah 711202, West Bengal, India}
\author{D. T.  Adroja} 
\email{devashibhai.adroja@stfc.ac.uk}
\affiliation{ISIS Facility, Rutherford Appleton Laboratory, Chilton, Didcot Oxon, OX11 0QX, United Kingdom} 
\affiliation{Highly Correlated Matter Research Group, Physics Department, University of Johannesburg, PO Box 524, Auckland Park 2006, South Africa}
\author{N. Kase}
\affiliation{Department of Applied Physics, Tokyo University of Science, Tokyo 125-8585, Japan}
\author{A. D. Hillier}
\affiliation{ISIS Facility, Rutherford Appleton Laboratory, Chilton, Didcot Oxon, OX11 0QX, United Kingdom} 
\author{A. M. Strydom} 
\affiliation{Highly Correlated Matter Research Group, Physics Department, University of Johannesburg, PO Box 524, Auckland Park 2006, South Africa}
\affiliation{Max Planck Institute for Chemical Physics of Solids, D-01187 Dresden, Germany}
\author{J. Akimitsu} 
\affiliation{Research Institute for Interdisciplinary Science, Okayama University,  Okayama, 700-8530 Japan}

\date{\today}

\begin{abstract}

We have examined the superconducting ground state properties of the caged type compound Sc$_5$Rh$_6$Sn$_{18}$ using magnetization, heat capacity, and muon-spin relaxation or rotation ($\mu$SR) measurements. Magnetization measurements indicate type-II superconductivity with an upper critical field $\mu_0H_{c2}(0)$ = 7.24 T. The zero-field cooled  and field cooled susceptibility measurements unveil an onset of diamagnetic signal below $T_{\bf c}$ = 4.4 K.  The interpretation of the heat capacity results below $T_{\bf c}$ using the $\alpha-$BCS model unveils the value of $\alpha$ = 2.65, which gives the dimensionless ratio 2$\Delta(0)/k_B T_{\bf c}$ = 5.3,  intimating that Sc$_5$Rh$_6$Sn$_{18}$ is a strong-coupling  BCS superconductor. The zero-field $\mu$SR measurements in the longitudinal geometry exhibit a signature of a spontaneous appearance of the internal magnetic field below the superconducting transition temperature, indicating that the superconducting state is characterized by the broken time-reversal symmetry (TRS). We have compared the results of broken TRS  in Sc$_5$Rh$_6$Sn$_{18}$ with that observed in R$_5$Rh$_6$Sn$_{18}$ (R = Lu and Y).
\end{abstract}
\pacs{71.20.Be, 75.10.Lp, 75.40.Cx}

\maketitle

Unconventional behaviour of superconductors beyond the conventional BCS theory is a major focus area in theoretical and experimental communities in condensed matter physics~\cite{Bardeen,Sigrist}.~BCS superconductors expel magnetic field through the Meissner effect. It is a very rare phenomenon for the superconducting ground state to support an internal magnetic field, which breaks the time reversal symmetry (TRS). TRS broken states were previously suggested for the high-temperature superconductors~\cite{htcref}, but their identification remains experimentally debatable. A symmetry breaking field can modify the superconducting ground state properties and may result in novel unconventional superconductivity\cite{Schaibley}. TRS breaking is rare and has only been observed directly in a few unconventional superconductors, e.g., Sr$_2$RuO$_4$~\cite{gm,jx}, UPt$_3$~\cite{gml}, (U;Th)Be$_{13}$~\cite{rhh}, (Pr;La)(Os;Ru)$_4$Sb$_{12}$~\cite{ya}, PrPt$_4$Ge$_{12}$~\cite{am}, LaNiC$_2$~\cite{ad1}, LaNiGa$_2$~\cite{ad2} and Re$_6$Zr~\cite{rps}.  The presence of an internal magnetic field places limitations on the pairing symmetry as well as on the possible mechanism responsible for superconductivity.

\par

In recent years, cage type compounds such as filled skutterudites (RT$_4$X$_{12}$) \cite{xs} where $R$ can be a rare-earth metal, $\beta-$pyrochlore oxides (AOs$_2$O$_6$) \cite{zh} where $A$ is an alkali metal, and Ge- or Si filled clathrates \cite{xy} have received much attention due to interesting aspects of the crystal structure that impedes heat conductivity in a manner that is considered to be beneficial to the design of novel thermoelectric materials. From a different point of view, a small number of so-called rattling materials among the cage-type structures also belong to the class of strongly correlated electron systems, and these are known for a rich variety of physics such as heavy fermion behavior, metal-insulator transition, multipole ordering, and superconductivity. RT$_4$X$_{12}$ and RT$_{2}$X$_{20}$ exhibit a strong interplay between quadrupole moment and superconductivity~\cite{kk,to,Onimaru}. Zero-field muon spin relaxation (ZF$-\mu$SR) is a powerful tool to search for TRS breaking fields or spontaneous internal magnetic fields below $T_{\bf c}$. The ZF-$\mu$SR measurements in PrOs$_4$Sb$_{12}$ (which was claimed to be the first Pr-based heavy fermion superconductor \cite{Bauer}) have revealed an appreciable increase in the internal magnetic field below the onset of superconductivity ($T_{\bf c}~=1.82~$K)~\cite{Koga}. The low-lying crystal-field excitations of Pr ions may be playing a vital role in the superconductivity~\cite{Koga}. The caged type material PrV$_2$Al$_{20}$ is a rare example of a heavy-fermion superconductor based on strong hybridization between conduction electrons and nonmagnetic quadrupolar moments of the cubic $\Gamma_3$ ground doublet. PrV$_2$Al$_{20}$ exhibits superconductivity at $T_{\bf c}=50~$mK in the antiferroquadrupole-ordered state under ambient pressure~\cite{Tsujimoto}. In the ordered state, the electronic heat capacity $C_e$ shows a $T^3$ temperature dependence, indicating the gapless mode associated with quadrupole order, octupole order, or both. PrIr$_2$Zn$_{20}$ and PrRh$_2$Zn$_{20}$ compounds exhibit non-Fermi liquid behavior in their resistivity and heat capacity and quadrupole ordering at low temperatures~\cite{Onimaru1}.  

\par
$R_5$Rh$_6$Sn$_{18}$ ($R=$ Sc, Y, Lu) compounds, having the caged type crystal structure also exhibit superconductivity (SC)~\cite{jpr} below $T_{\bf c}$ = $4.4~$K (Sc), $3~$K (Y), and $4~$K (Lu). These compounds have a tetragonal structure with the space group $I4_1/acd$ and rare-earth element coordination $Z=8$, and where $R$ occupies two different crystallographic sites ~\cite{sm}. The crystal structure is similar to the skutterudite structure~\cite{Bauer}.~Lu$_5$Rh$_6$Sn$_{18}$ is a conventional BCS type superconductor~\cite{kase2}. The gap structure of Y$_5$Rh$_6$Sn$_{18}$ is found to be strongly anisotropic as revealed from the heat capacity measurements; $C_e(T)$ exhibits a $T^3$ variation and $C_P(H)$, where $H$ is the applied magnetic field indicates a $\sqrt{H}$-like dependence~\cite{kase2}. The superconducting properties of Y$_5$Rh$_6$Sn$_{18}$ thus have a similarity with those of the anisotropic $s$-wave superconductor YNi$_2$B$_2$C except for the difference in $T_{\bf c}$~\cite{kase2}. Zero field, transverse field and longitudinal field muon spin relaxation measurements on Y$_5$Rh$_6$Sn$_{18}$ have been reported by our group~\cite{Bhattacharyya}. For Lu and Y compounds, the resistivity $\rho(T)$ exhibits an unusual temperature variation. In the Lu compound $\rho(T)$ is nearly constant down to $120~$K, and shows an increase on further cooling. For the Y compound $\rho$ continuously increases on cooling below room temperature, with a kink appearing at about $120 ~$K. Coexistence of superconductivity and magnetism was observed in the Tm$-$based reentrant superconductor Tm$_5$Rh$_6$Sn$_{18}$ ($T_{\bf c}$ = 2.2 K)~\cite{rojek, kase3}. 

\par
We have recently reported superconducting properties of the caged type compounds (Lu,Y)$_5$Rh$_6$Sn$_{18}$ using magnetization, heat capacity, and muon-spin relaxation ($\mu$SR) measurements~\cite{Bhattacharyya,Bhattacharyya1}. Zero-field $\mu$SR measurements reveal the spontaneous appearance of an internal magnetic field below the superconducting transition temperature, which indicates that the superconducting state in these materials is characterized by broken time-reversal symmetry~\cite{Bhattacharyya1}.  It is interesting to note that the electronic heat capacity ($C_e$) of Lu$_5$Rh$_6$Sn$_{18}$ exhibits exponential behavior as a function of temperature below $T_{\bf c}$~\cite{nk1,nk2}. From a series of experiments on $R_5$Rh$_6$Sn$_{18}$ ($R=$Lu, Sc, Y and Tm), it was concluded that the gap structure is strongly dependent on the $R$ atom, whose origin is left to be clarified~\cite{rojek, kase3}. In this Rapid Communication, we address these matters by ZF$-\mu$SR measurements for the Sc$_{5}$Rh$_{6}$Sn$_{18}$ system. The results unambiguously reveal the spontaneous appearance of an internal magnetic field in the SC state, providing clear evidence for broken time reversal symmetry and suggesting a common origin in this family of compounds.

\begin{figure}[t]
%\vskip -1cm
\centering
\includegraphics[width=\linewidth]{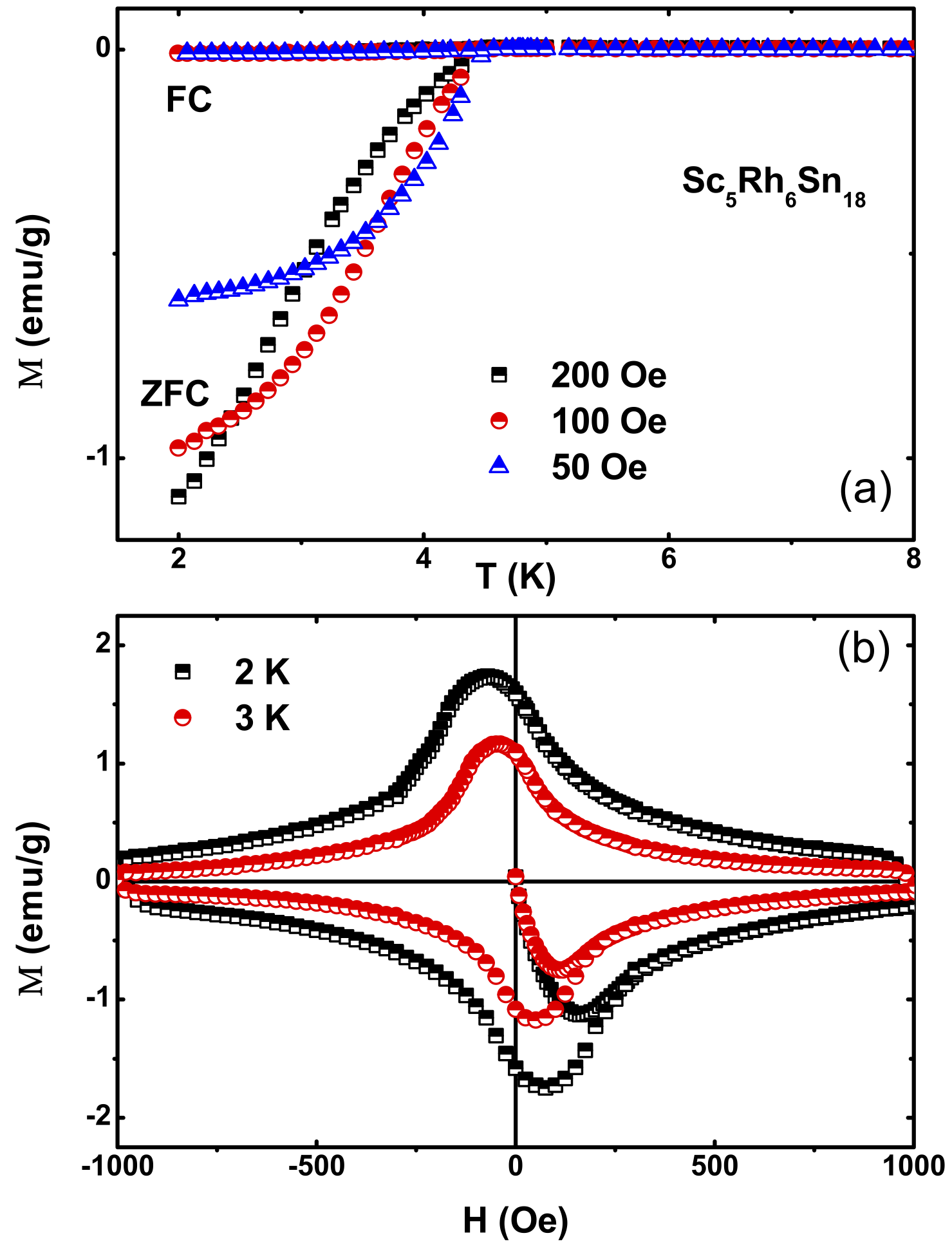}
\vskip 0.5cm
\caption {(Color online) (a) Temperature dependence of the magnetic susceptibility of Sc$_{5}$Rh$_{6}$Sn$_{18}$ under magnetic fields of 50, 100 and $200~$G in the zero-field cooled (ZFC) and field cooled (FC) states.~(b) Isothermal field dependence of magnetization at $2$ and $3~$K.}
\end{figure}

The single crystals of Sc$_5$Rh$_6$Sn$_{18}$ were grown by dissolving the constituent elements in an excess of Sn-flux in the ratio of Sc:Rh:Sn = 1:2:20. The quartz tube was heated up to 1050$^\circ$C, maintained at this temperature for about $3~$h, and cooled down to $200^\circ$C at a rate of $5^\circ$C/h, taking $7$ days in total. The excess flux was removed from the crystals by spinning the ampoule in a centrifuge~\cite{jpr}.~Laue patterns were recorded using a Huber Laue diffractometer and well defined Laue diffraction spots indicate the high quality of the single crystals. The phase purity was inferred from the powder x-ray  patterns which were indexed as the Sc$_5$Rh$_6$Sn$_{18}$ phase with the space group~\cite{jpr} $I4_1/acd$. The magnetization data were collected using a Quantum Design Superconducting Quantum Interference Device. The heat capacity measurements were performed down to 500 mK using a Quantum Design Physical Properties Measurement System equipped with a $^3$He refrigerator.

\par

We further employed the $\mu$SR technique to investigate the superconducting ground state. The $\mu$SR measurements were performed at the MUSR spectrometer at the ISIS Neutron and Muon Facility located at the STFC Rutherford Appleton Laboratory (RAL, United Kingdom). The single crystals (typical size 3$\times$3$\times$3 mm$^3$) were mounted on a high purity silver plate (99.995\% silver) using diluted GE varnish and then cooled down to $1.2~$K in a standard $^4$He cryostat with He-exchange gas. It is to be noted that due to the small size and irregular shape the crystals were not aligned in a  particular direction, but had random orientations with respect to the incident muon beam. Using an active compensation system the stray magnetic fields at the sample position were canceled to a level of 1 mG. Spin-polarized muons were implanted into the sample and the positrons from the resulting muon decay were collected in the detector positions either forward or backwards of the initial muon spin direction. The asymmetry of the muon decay is calculated by; $G_z(t) =[ {N_F(t) -\alpha N_B(t)}]/[{N_F(t)+\alpha N_B(t)}]$, where $N_B(t)$ and $N_F(t)$ are the number of counts at the detectors in the forward and backward positions and $\alpha$ is a constant determined from calibration measurements made in the normal state with a small $20~$G  transverse applied magnetic field. The data were analyzed using the software package Wimda~\cite{FPW}.  

\par

The bulk nature of superconductivity in Sc$_5$Rh$_6$Sn$_{18}$ was confirmed by the magnetic susceptibility $\chi(T)$, as shown in Fig.\ 1(a). The low-field $\chi(T)$ measurements display a strong diamagnetic signal due to the superconducting transition at  $T_{\bf c}=4.4~$K. Fig.\ 1(b) shows the magnetization $M(H)$ at $2~$K and at $3~$K with a shape that is typical for type-II superconductivity.  The electrical resistivity (not shown) exhibits bulk superconductivity at $4.4~$K~\cite{nk1,nk2}. 

\par

Fig.\ 2(a) shows  $C_P(T)/T$ vs. $T^2$  in field values $H=0$ and $7.5~$T. At $4.4~$K a sharp anomaly is observed indicating the superconducting transition which matches well with $\chi(T)$ data. Since the normal-state heat capacity was found to be invariant under external magnetic fields, the normal-state electronic heat capacity coefficient $\gamma$ and the lattice heat capacity coefficient $\beta$ were deduced from the data in a field of $7.5~$T where the superconductivity is completely suppressed, using a least-square fit of the $C_P(T)/T$ data to $C_P(T)/T = \gamma +\beta T^2 + \delta T^4$. The least-squares analysis of the $7.5~$T data provides a Sommerfeld constant $\gamma$ = 51.10 mJ/(mol-K$^2$), $\beta$ = 0.13 mJ/(mol-K$^4$), $\delta$ = 0.32 mJ/(mol-K$^6$) and from this value of  $\beta$ we have estimated the Debye temperature $\Theta_D=271~$K~\cite{nk1,nk2}. We have analyzed the electronic heat capacity data (below $T_{\bf c}$) using $T^3$ model and the single-band $\alpha-$model that was adapted from the single-band BCS theory to fit the heat capacity data that deviate from the BCS prediction~\cite{cp1,cp2}. The red and blue solid lines in Fig.\ 2(b) demonstrate a theoretical fit based upon the $\alpha-$ model and $T^3$ model. In the $\alpha-$ model it was assumed that the normalized gap amplitude $\Delta(T)/\Delta(0)$ follows the isotropic $s$-wave BCS result with $\alpha$ =  $\Delta(0)/k_B T_{\bf c}$ being an adjustable parameter~\cite{nk1,nk2}. The $\alpha$-model is an excellent fit to the electronic heat capacity data of Sc$_5$Rh$_6$Sn$_{18}$ below $T_{\bf c}$ with $\alpha = 2.65$, which is significantly larger than the value for the weak-coupling BCS value of 1.76. All of these results suggest that Sc$_5$Rh$_6$Sn$_{18}$ is a strong-coupling superconductor with the value of $2\Delta(0)/k_BT_{\bf c}$ = 5.3.  The Ginzburg-Landau (GL) coherence length $\xi(0)$ and the GL parameter $\kappa(0)$ = $\lambda(0)/\xi(0)$ can be obtained from the upper critical field $H_{c2}{(0)}$, the lower critical field $H_{c1}{(0)}$, and $H_c(0)$ using the following equations: $\mu_0 H_{c2}{(0)}$ = $\Phi_0$/2$\pi \xi(0)^2$, $H_{c1}(0)$ = $H_c(0)^2/H_{c2}(0)[\ln\kappa(0) +0.08]$, $H_c(0)$ = $H_{c2}(0)/\sqrt{2}\kappa(0)$. From these, $\lambda(0)$ and $\xi(0)$ are estimated to be approximately 34.2 nm and 6.74 nm, respectively. In addition, $\kappa(0)$ is calculated to be 51.7. Because $\kappa$ is larger than 1/$\sqrt{2}$, Sc$_5$Rh$_6$Sn$_{18}$ is a type II superconductor.

\begin{figure}[t]
\centering
\includegraphics[width =\linewidth]{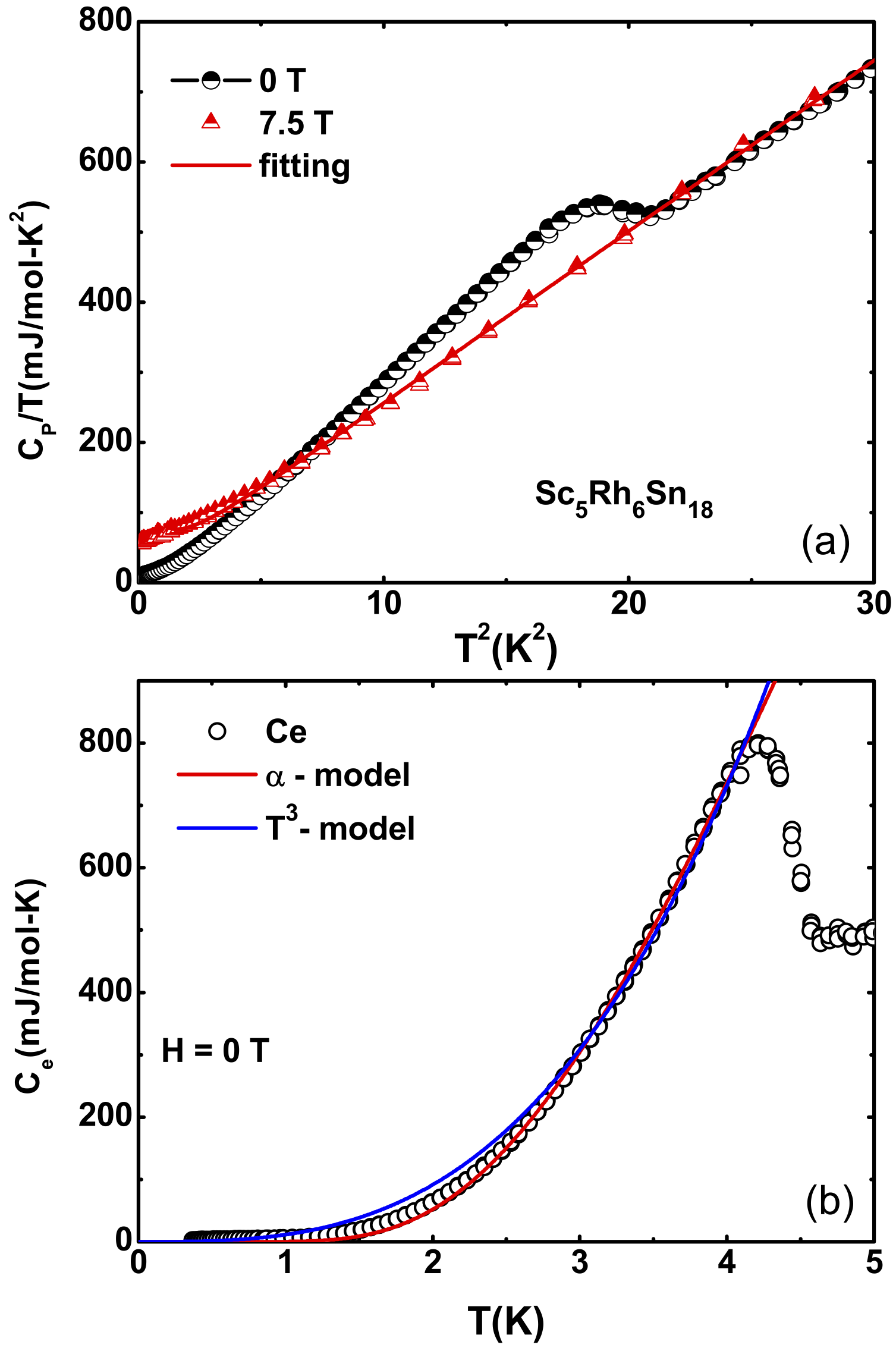}
\caption {(Color online) (a) $C_P/T$ {\it vs}.\ $T^2$ in two different applied magnetic field values. The solid line shows a fit to the $H=7.5~$T data (see text) where $T_{\bf c}$ is suppressed to far below the shown temperature range of measurement. (b) Temperature dependence of electronic heat capacity $C_e$ under zero field after subtracting the lattice contribution for Sc$_5$Rh$_6$Sn$_{18}$.}
\end{figure}

\begin{figure}[t]
\centering
\includegraphics[width=\linewidth]{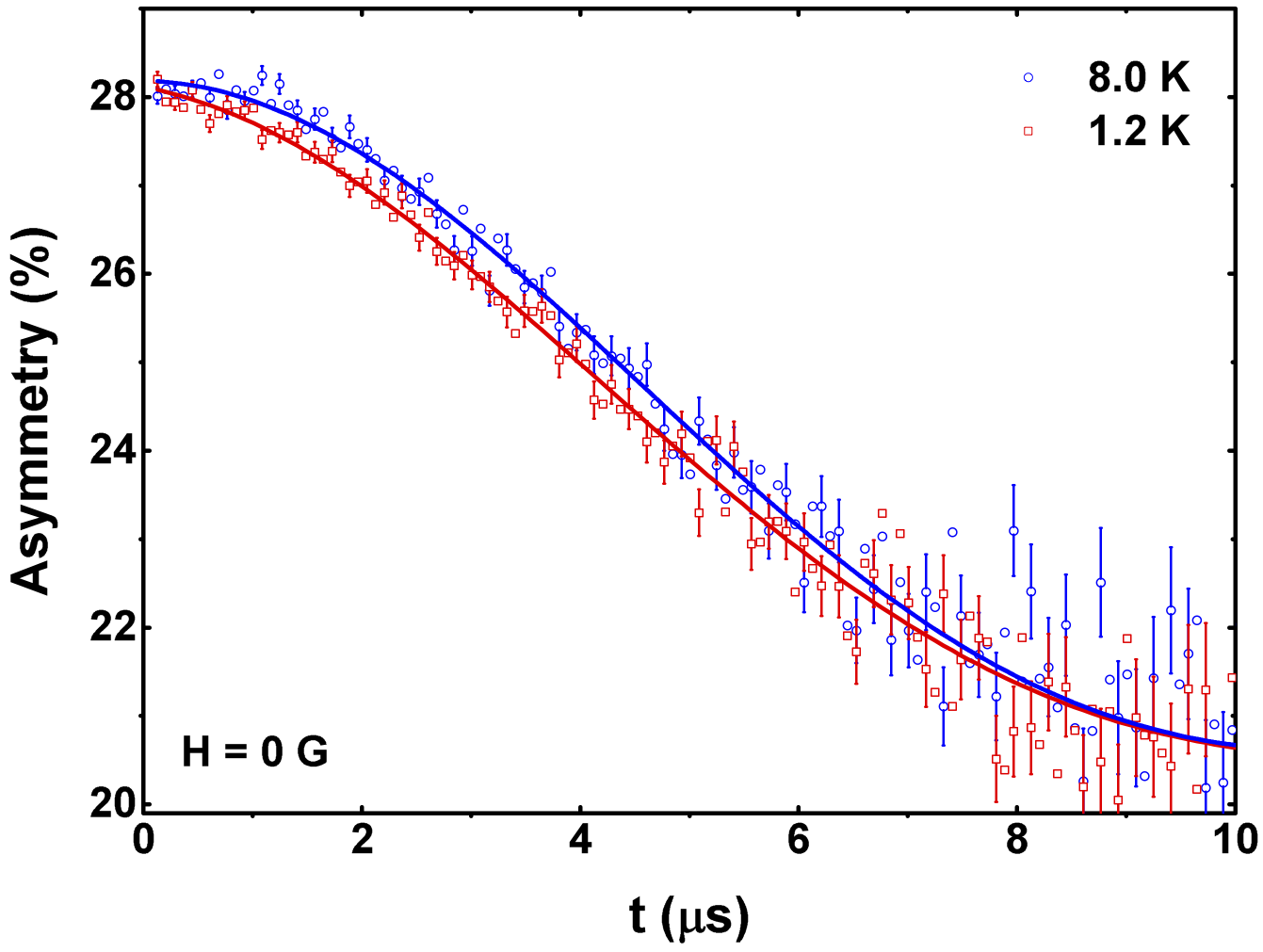}
\caption {(Color online) Zero-field $\mu$SR time spectra for Sc$_5$Rh$_6$Sn$_{18}$ collected at $1.2~$K (squares) and $8.0~$K (circles) are shown together with lines that are least-squares fits to the data using Eq.\ (1). These spectra collected below and above $T_{\bf c}$ are representative of the data collected over a range of temperatures.}
\end{figure}

\par
Fig.\ 3 shows the time evolution of the zero field muon spin relaxation asymmetry in Sc$_5$Rh$_6$Sn$_{18}$ at temperatures above and below $T_{\bf c}$. Below $T_{\bf c}$, we observed that the muon spin relaxation became faster with decreasing temperature down to lowest temperature, which indicates the appearance of a spontaneous magnetic field in the superconducting phase. We note that there is no signature of muon spin precession that would accompany a sufficiently large internal magnetic field produced by ordering of electronic moments. The ZF$-\mu$SR spectra for Sc$_5$Rh$_6$Sn$_{18}$ can be well described by the damped Gaussian Kubo-Toyabe (K-T) function~\cite{Adroja1,Adroja2,Adroja3,Bhattacharyya3},

\begin{equation}
G_{z2}(t) =A_0 G_{KT}(t)e^{-\lambda t}+A_{bg}~,
\end{equation}
where
\begin{equation}
G_{KT}(t) =\left[\frac{1}{3}+\frac{2}{3}(1-\sigma_{KT}^2t^2)e^{{\frac{-\sigma_{KT}^2t^2}{2}}}\right]
\end{equation}

is the K-T functional form expected from an isotropic Gaussian distribution of randomly oriented static (or quasi-static) local fields at muon sites. $\lambda$ is the electronic relaxation rate, $A_0$ is the initial asymmetry, and $A_{bg}$ is the background arising from the muons stopping on the silver sample holder. $A_0$ and $A_{bg}$ are all found to be temperature independent. First we estimated the value of Kubo-Toyabe depolarization rate $\sigma_{KT}$ by fitting the data at the lowest temperature and then kept this value fixed for fitting other temperature data points as shown in Fig.\ 4(b) for ZF$-\mu$SR fitting as there is negligible variation with temperature in $\sigma_{KT}$ within the error bars. 

\par

Fig.\ 4(a) shows the temperature dependence of the electronic relaxation rate. It is remarkable that $\lambda$ shows a significant increase below an onset temperature of $4.4\pm 0.1~$K , but $\sigma_{KT}$ is temperature independent (see Fig. 4 (b)), indicating the appearance of a spontaneous internal field correlated with the superconductivity.  This observation provides unambiguous evidence that TRS is broken in the SC state of Sc$_5$Rh$_6$Sn$_{18}$. Such a change in $\lambda$ has only been observed in superconducting URu$_2$Si$_2$~\cite{Schemm}, Sr$_2$RuO$_4$~\cite{gm}, LaNiC$_2$~\cite{ad1}, (Lu, Y)$_5$Rh$_6$Sn$_{18}$~\cite{Bhattacharyya1} and SrPtAs~\cite{pkb}.  This increase in $\lambda$ can be explained by the presence of a very small internal field as discussed by Luke {\it et al.\ }~\cite{gm}, for Sr$_2$RuO$_4$. This suggests that the field distribution is Lorentzian in nature similar to the case of Sr$_2$RuO$_4$. Considering a similar temperature dependence of $\lambda$ in Sr$_2$RuO$_4$, LaNiC$_2$, SrPtAs, Lu$_5$Rh$_6$Sn$_{18}$ and  Sc$_5$Rh$_6$Sn$_{18}$, we attribute this behavior of $\lambda$ to TRS breaking below $T_{\bf c}$ in Sc$_5$Rh$_6$Sn$_{18}$. It is to be noted that for Y$_5$Rh$_6$Sn$_{18}$ the onset of a TRS breaking~\cite{Bhattacharyya1} field appears in $\lambda(T)$ below $2~$K which is well below $T_{\bf c}=3.0~$K. The increase in the exponential relaxation below $T_{\bf c}$ in Sc$_5$Rh$_6$Sn$_{18}$ is $0.0214~\mu$s$^{-1}$, which corresponds to a characteristic field strength $\lambda/\gamma_\mu=0.25~$G, where $\gamma_\mu$ is the muon gyromagnetic ratio equal to $2\pi\times 135.5~$MHz/T. This is about half the value that was observed in the case of Lu$_5$Rh$_6$Sn$_{18}$, in the B phase of UPt$_3$ and in Sr$_2$RuO$_4$~\cite{gml}. No theoretical estimates of the characteristic field strength in Sc$_5$Rh$_6$Sn$_{18}$ are yet available; however, we expect it to be comparable to those in Sr$_2$RuO$_4$ and UPt$_3$ as the fields are expected to arise from a similar mechanism. On the other hand the TRS breaking field appears in $\sigma_{KT}(T)$ in LaNiGa$_2$~\cite{laniga2} and in PrOs$_4$Sb$_{12}$~\cite{prossb}.

\begin{table}[t]
\begin{center}
\caption{Comparison of superconducting parameters of R$_5$Rh$_6$Sn$_{18}$ [R = Sc, Lu and Y] compounds\cite{Bhattacharyya,Bhattacharyya1} with other TRS breaking  superconductors~\cite{AOKI,MacLaughlin, Suderow}.\\}
\begin{tabular}{lccccccccccccccc}
\hline
\hline
 Compounds && $H_{c2}$ && $T_{\bf c}$ && Gap to $T_{\bf c}$ ratio &&  $\Delta\lambda$   && Gap  \\ 
&& (T)  && (K) && 2$\Delta(0)/k_B T_{\bf c}$ &&    (G) &&  function  \\ 
\hline
Sc$_5$Rh$_6$Sn$_{18}$  && 7.24 && 4.4 && 5.3 &&  0.6 && $s-$wave$^{\rm a}$  \\ 
Lu$_5$Rh$_6$Sn$_{18}$ && 6.45 &&  4.0 && 4.28 &&  0.5 &&  $s-$wave~\cite{Bhattacharyya2} \\ 
Y$_5$Rh$_6$Sn$_{18}$ && 3.13 && 3.0 && 4.23 &&  0.02  &&  $s-$wave~\cite{Bhattacharyya1} \\ 
PrOs$_4$Sb$_{12}$ && 2.2 && 1.8 && 3.7 &&  1.2 && s/p-wave~\cite{MacLaughlin}\\ 
 &&  &&  &&  &&   							&& or s-wave ~\cite{Suderow,AOKI}\\ 
UPt$_3$ && 1.9 && 0.52 && 2.0  &&  0.1  && $p / f-$wave~\cite{UPt3Broholm}\\ 
Sr$_2$RuO$_{4}$ &&  && 1.5 && 3.5 &&  0.5 && $d-$wave~\cite{SrFirmo} \\ 
\hline 
\footnotetext{from heat capacity analysis}
\end{tabular}
\end{center}
\end{table}

\begin{figure}[t]
\vskip -0.5 cm
\centering
\includegraphics[width=\linewidth]{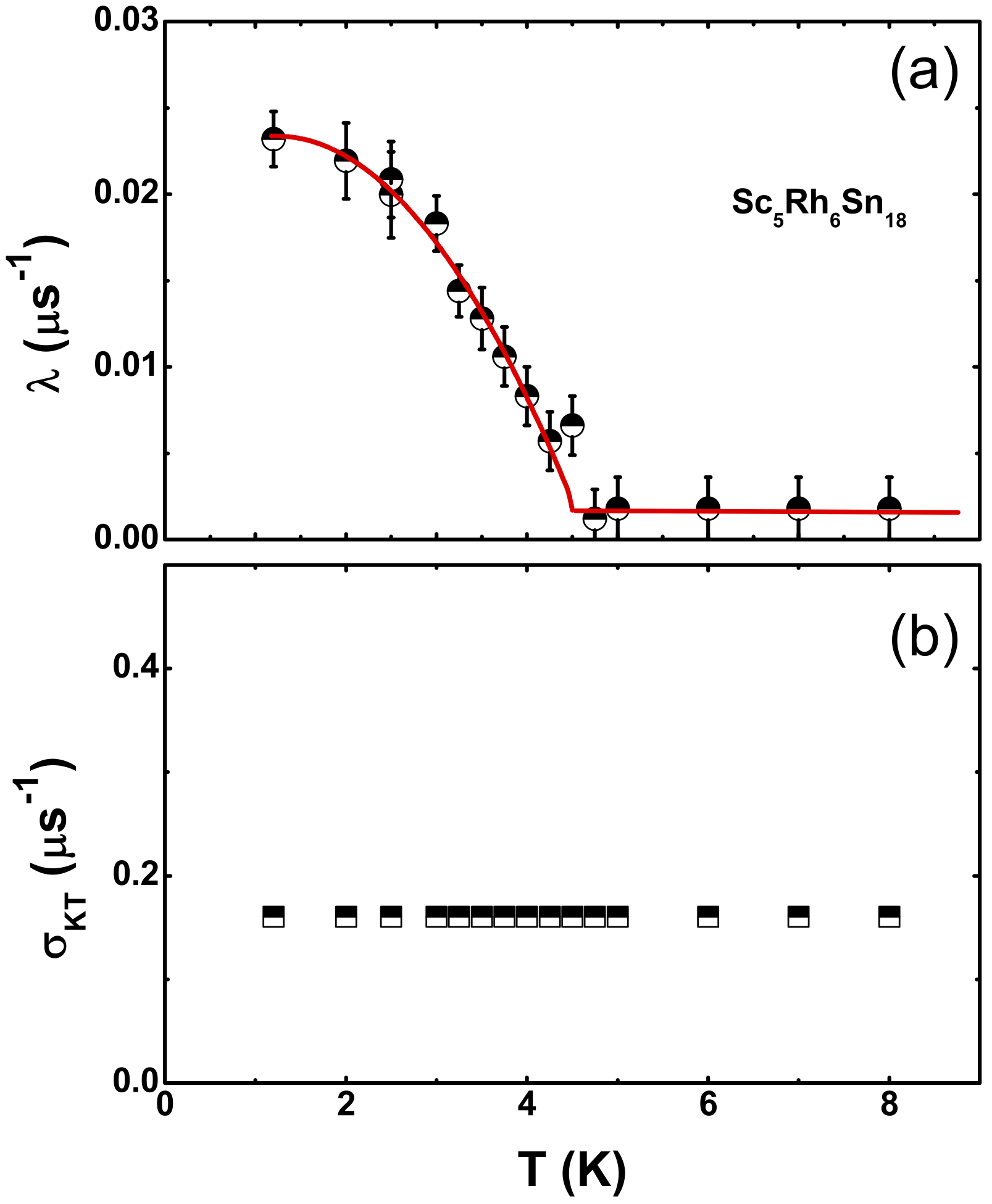}
\caption {(Color online) (a) Temperature dependence of the electronic relaxation rate measured in zero magnetic field of Sc$_5$Rh$_6$Sn$_{18}$ with $T_{\bf c}=4.4~$K. The lines are guides to the eye. The relaxation is near zero above $T_{\bf c}$, but rises decisively from right at $T_{\bf c}$ which indicates the presence of an internal magnetic field and, consequently, suggests the superconducting state has broken time reversal symmetry. (b) The Kubo-Toyabe  depolarization rate $\sigma_{KT}$, versus temperature in zero field shows no temperature dependence.}
\end{figure}

\par

Our theoretical analysis~\cite{Sigrist,Mazidian,Bhattacharyya2} for the isostructural compound Lu$_5$Rh$_6$Sn$_{18}$ was carried out under the assumption of strong spin orbit coupling and revealed two possible superconducting pairing states. The first one has singlet $d+id$ character and the second one has triplet non-unitarity character. Far below the superconducting temperature $T \ll T_{\bf c}$, the thermodynamics of the singlet state would be influenced by a line node, which suggest a quadratic temperature dependence of the heat capacity. Furthermore, the triplet state~\cite{Mazidian} would be influenced by point nodes, which happen to be shallow (a result protected by symmetry) and therefore also lead to quadratic temperature variation of the heat capacity. Nevertheless, because of the location of the nodes in the triplet case, fully-gapped behavior may be recovered depending on the topology of the Fermi surface. In addition some limiting cases of the triplet state correspond to regular, {\it i.e.} linear point nodes (cubic temperature dependence of the heat capacity) as well as to a more exotic state with a nodal surface (gapless superconductivity, linear temperature variation of the heat capacity). We note that the theoretical analysis presented in the supplemental material~\cite{Bhattacharyya2} in Ref [48] is valid for any superconductor with $D_{4h}$ point group symmetry in the presence of strong spin-orbit coupling and broken time-reversal symmetry and may therefore be applied for example to Sr$_2$RuO$_4$~\cite{Veenstra}, as well as to Sc$_5$Rh$_6$Sn$_{18}$.

\par

In summary, we have investigated the nature of the superconducting ground state in Sc$_5$Rh$_6$Sn$_{18}$ by using ZF$-$$\mu$SR measurements. Below $T_{\bf c}=4.4~$K, the ZF$-$$\mu$SR measurements revealed the onset of an appreciable internal magnetic field that is correlated exactly with the onset of superconductivity in this compound. The appearance of spontaneous magnetic fields in our ZF-$\mu$SR results provide unambiguous evidence for TRS breaking in this material and an unconventional pairing mechanism. The evidence of broken TRS in the SC state will help to narrow down the number of possibilities for the symmetry of the SC order parameter. Symmetry analysis suggests either a singlet $d+id$ state with a line node or, alternatively, non-unitary  triplet pairing with point nodes, which may be linear or shallow and can become fully gapped depending on the Fermi surface topology. It is hoped that our experimental results presented in this paper will stimulate theoretical interest to understand the unconventional superconductivity in cage type superconductors, as well as to understand the origin of a TRS breaking field in either the electronic/$\lambda (T)$ or the nuclear/$\sigma_{KT}(T)$ channel. 

\section*{}

We would like to thank J. Quintanilla for interesting theoretical discussions. AB would like to acknowledge DST India, for Inspire Faculty Research Grant, FRC of UJ, NRF of South Africa and ISIS-STFC for funding support. DTA and ADH would like to thank CMPC-STFC, grant number CMPC-09108, and also DIST for financial support. DTA thanks to JSPS for invitation fellowship. AMS thanks the SA-NRF (Grant 93549) and UJ Research Committee for financial support.

\end{document}